\documentclass{article}

\newcommand{\dddot}[1]{\stackrel{\mathbf{\ldots }}{#1}}
\begin{document}

\title{\LARGE \bf The high energy limit of the trajectory \\
representation of quantum mechanics}

\author{Edward R.\ Floyd \\
10 Jamaica Village Road, Coronado, CA 92118-3208, USA \\
floyd@mailaps.org}

\date{ }

\maketitle

\begin{abstract}
The trajectory representation in the high energy limit (Bohr correspondence 
principle) manifests a residual indeterminacy.  This indeterminacy is compared 
to the indeterminacy found in the classical limit $(\hbar \to 0)$ [Int. J. Mod. 
Phys. {\bf A 15}, 1363 (2000)] for particles in the classically allowed region, 
the classically forbidden region, and near the WKB turning point.  The 
differences between Bohr's and Planck's principles for the trajectory 
representation are compared with the differences between these correspondence 
principles for the wave representation.  The trajectory representation in the 
high energy limit is shown to go to neither classical nor statistical mechanics.  
The residual indeterminacy is contrasted to Heisenberg uncertainty.  The 
relationship between indeterminacy and `t~Hooft's information loss and 
equivalence classes is investigated. 
\end{abstract}

\bigskip

\footnotesize
\noindent PACS Numbers: 3.65.Bz; 3.65.Ca

\noindent Key words: high energy limit, trajectory interpretation,
Bohr's correspondence principle, residual indeterminacy,
't~Hooft's equivalence classes, Heisenberg uncertainty principle.
\normalsize

\section{INTRODUCTION} 
Liboff has studied the Bohr correspondence principle
(the high energy limit) of the Schr\"odinger representation of quantum 
mechanics.$^{\ref{bib:liboff}}$  Liboff found that the Bohr correspondence 
principle and the Planck correspondence principle ($\hbar \to 0$) are not 
equivalent.  Liboff, who worked in the Schr\"odinger representation, used 
frequency (spectral) correspondence and form correspondence (the probability 
density) to investigate the Bohr correspondence.  He found that neither the 
frequency nor the form correspondences apply to all systems.  While we shall not 
investigate the Schr\"odinger representation here, I do make one comment on one 
of Liboff's counterexamples.  Liboff studied the particle confined in a cubical 
box with impenetrable walls.$^{\ref{bib:liboff}}$  Impenetrable walls imply 
infinitely jumps (discontinuities) in the potential.  The particle in this box 
does not permit a short wave limit because it always has an infinitely large 
relative change of wavelength at the impenetrable wall.$^{\ref{bib:ajp60}}$  
Such a system violates Bohr correspondence by design.

Recently, Planck correspondence ($\hbar \to 0$) has been studied for the 
trajectory representation.$^{\ref{bib:f2000}}$  Herein we investigate Bohr 
correspondence for the trajectory representation of quantum mechanics.  Bohr 
correspondence purports that nonrelativistic quantum mechanics reduces to 
nonrelativistic classical mechanics in the high energy (often called large $n$) 
limit.  Of course, this is an idealization that ignores the applicability of 
relativity in the high energy limit.  Following the precedent of 
Bohr$^{\ref{bib:bohr},\ref{bib:rosenfeld}}$ and Liboff$^{\ref{bib:liboff}}$, we 
also prosecute our investigation couched in the nonrelativistic trajectory 
representation assuming that the character of the limit is apparent before 
relativistic corrections are needed.  The present state of development of a 
relatistic trajectory representation is 
modest.$^{\ref{bib:f1988},\ref{bib:bfm}}$ 
   
Recently, Faraggi and Matone have shown that, while quantum mechanics is 
compliant with the equivalence principle where all quantum systems can be
connected by an equivalence coordinate transformation (trivializing
map) given by $x \to \tilde{x}(x)$, classical mechanics is not.$^{\ref{bib:fm1}-
\ref{bib:fm3}}$  Note that this coordinate transformation is much more stringent 
than contact and canonical transformations.  Faraggi and Matone through the 
equivalence principle have
independently derived, free from axioms, the quantum  
Hamilton-Jacobi foundation for the trajectory representation of
quantum mechanics.  This frees the trajectory representation from all Copenhagen 
philosophy that has been imposed upon the Schr\"odinger representation of 
quantum mechanics.  We theoretically test Bohr's
correspondence principle$^{\ref{bib:bohr},\ref{bib:rosenfeld}}$ by investigating 
the transition from quantum to classical mechanics for the trajectory
representation, embedded in a non-relativistic quantum 
Hamilton-Jacobi theory, for the energy growing without bound.

We examine the high energy limit for three cases: a particle in the
classically allowed region; a particle in the classically forbidden
region (beyond the WKB turning point); and a particle in the
vicinity of the WKB turning point.  We choose potentials that are
heuristic and whose trajectory representation can be presented in closed form by
familiar functions.  One dimension suffices for this investigation.

The quantum stationary Hamilton-Jacobi equation (QSHJE) is a
phenomenological equation just like the classical stationary
Hamilton-Jacobi equation (CSHJE) and the Schr\"{o}dinger equation. 
The QSHJE and CSHJE render generators of the motion that, for the same 
potential, describe trajectories that differ.  As we shall show herein, these
trajectories still generally differ in the high energy limit.  The QSHJE in one
dimension, $x$, is given for non-relativistic quantum motion
by$^{\ref{bib:fm1},\ref{bib:messiah},\ref{bib:prd34}}$ 

\begin{equation}
\frac{W_x^2}{2m}+V-E=-\frac{\hbar^2}{4m}\langle W;x\rangle
\label{eq:qshje}
\end{equation}

\noindent where $W$ is the reduced action (also known as Hamilton's
characteristic function), $W_x$ is the conjugate momentum, $V$ is
the potential, $\hbar=h/(2\pi )$, $h$ is the Planck constant, $m$
is the mass, and $\langle W;x\rangle$ is the Schwarzian derivative
that manifests the quantum effects.  The Schwarzian derivative
contains higher-order derivative terms given by

\[
\langle W;x\rangle = \frac{W_{xxx}}{W_x} - \frac{3}{2} \left(
\frac{W_{xx}}{W_x}\right)^2.
\]

\noindent  The complete solutions for the reduced action and
conjugate momentum are well known and given for energy $E$
by$^{\ref{bib:prd34},\ref{bib:afb20}}$

\begin{equation}
W = \hbar \arctan \left(\frac{b \theta /\phi + c/2}{(ab-
c^2/4)^{1/2}}\right) + K,\ \ \ \ x>0.
\label{eq:ra}
\end{equation} 

\noindent and

\begin{equation}
W_x=(2m)^{1/2}(a\phi ^2 + b\theta ^2 + c\phi \theta)^{-1}
\label{eq:cme}
\end{equation}

\noindent where $K$ is an integration constant that we may
arbitrarily set $K=0$ for the rest of this investigation and where
$(\phi ,\theta)$ is the set of independent solutions to the
associated stationary Schr\"{o}dinger equation for energy $E$.  The
Wronskian ${\cal W}(\phi ,\theta)$ is normalized so that ${\cal
W}^2 = 2m/[\hbar ^2(ab-c^2/4)]$.  The set of coefficients $(a,b,c)$
for energy $E$ determines the particular solution for $W$ and
$W_x$.  The set $(a,b,c)$ is determined by a sufficient set of a
combination of initial values or constants of the motion other than
$E$ for the third-order QSHJE.  This requires three independent
values.  For example, the set of coefficients $(a,b,c)$ can be
specified by the initial values $[W_x(x_o),W_{xx}(x_o)]$ for the
QSHJE plus the Wronskian, ${\cal W}$, which is a constant for the
Schr\"{o}dinger equation.  The particular solution was shown to
specify the particular microstate which had not been detected in
the Schr\"{o}dinger representation.$^{\ref{bib:pla249},\ref{bib:prd34},
\ref{bib:carroll},\ref{bib:carroll2}}$  The left side of the QSHJE, Eq.\
(\ref{eq:qshje}), is the CSHJE.  If $\hbar $ were zero (as
distinguished from the limit $\hbar $ going to zero), then Eq.\
(\ref{eq:qshje}) would trivially reduce to the classical 
time-independent Hamilton-Jacobi equation.  Herein, we test
Bohr's correspondence principle by investigating the proposition
that quantum mechanics transitions to classical mechanics in the
limit that $E$ grows without bound.  The trajectory representation of
quantum mechanics is well suited for testing Bohr's
correspondence principle.  While the CSHJE is a first-order
nonlinear differential equation, the QSHJE is a third-order
nonlinear differential equation whose second- and third-order terms
appear in $\langle W;x \rangle$ as fractions where the numerators and 
corresponding denominators contain derivative terms of the same degree with 
regard to exponents.  This investigation must include, in the limit $E \to 
\infty$, the effects of annulling all higher than first order terms in a 
differential equation upon the particular solution and upon the set of necessary 
and sufficient initial values.  

With the trajectory representation, we show that quantum mechanics
in the high energy limit does not generally reduce to classical
mechanics.  Nor does it reduce to statistical mechanics even though a
residual indeterminacy exists in general in the high energy limit.  We show that 
in the classically allowed region Bohr correspondence and Planck correspondence 
for the trajectory representation are similar; in the neighborhood of the WKB 
turning point, partly similar; and in the classically forbidden region, not 
similar.  We investigate this residual indeterminacy and contrast it to 
Heisenberg uncertainty.  We also gain insight
into the quantum term $\hbar ^2\langle W;x \rangle /(4m)$.

Recently, 't~Hooft proposed that underlying contemporary quantum
mechanics there should exist a more fundamental, albeit still
unknown, theory at the Planck level that would provide more
information than the Schr\"{o}dinger wave
function.$^{\ref{bib:thooft}}$  In 't~Hooft's proposal, the
additional information distinguishes primordial states at the
Planck level, but this information is lost through dissipation as
these states evolve into states forming an equivalence class. 
't~Hooft suggested quantum gravity would dissipate information. 
States of an equivalence class, after a while, become
indistinguishable from each other even though they have different
pasts.  't~Hooft's ideas bear upon this investigation by giving us
insight into the relationship of the Copenhagen interpretation to
the trajectory representation where information regarding
microstates (primordial states) becomes ``lost" and where the
Schr\"{o}dinger wave function becomes an equivalence class.  We
compare residual indeterminacy with 't~Hooft's information loss and
equivalence classes in the high energy limit.  This comparison is preliminary 
because the underlying fundamental theory of 't~Hooft's proposal is not yet
complete.   

In Section 2, we investigate the Bohr limit of the trajectory
representation of quantum mechanics for a particle in the
classically allowed region.  In Section 3, we examine a trajectory
in the Bohr limit in the classical forbidden region.  We
investigate, in Section 4, a trajectory as it transits between the
classically allowed and forbidden regions across the WKB turning
point.  In Section 5, we discuss the impact of the Bohr limit
upon the set of initial values necessary and sufficient to specify
the trajectory.  In Section 6, we present the relationship between this 
investigation and 't~Hooft's information loss and
equivalence classes.  In Section 7, we contrast residual
indeterminacy to Heisenberg uncertainty. 

\section{THE CLASSICALLY ALLOWED CASE}  
Let us examine a particle in the classically allowed region, $E>V(x)$, in the 
limit that $E$ grows without bound.  In the limiting process as $E$ grows 
without bound, the quantum motion does not approach the classical conjugate 
motion even though the left hand side of the QSHJE grows without bound while the 
right side remains finite.  Quantum physics is more subtle.  

Let us consider a heuristic example.  We choose a free particle,
$V=0$.  An acceptable set of independent
solutions to the Schr\"{o}dinger equation for the free particle is
given by

\begin{equation}
\phi = [E(ab-c^2/4)]^{-1/4}\cos [(2mE)^{1/2}x/\hbar ] \ \ \ \
\mbox{and} \ \ \ \ \theta = [E(ab-c^2/4)]^{-1/4}\sin
[(2mE)^{1/2}x/\hbar].
\label{eq:isc}     
\end{equation}

\noindent The coefficients $(a,b,c)$ specify the particular
microstate for a specified energy $E$ in accordance with a
sufficient set of a combination of initial values and constants of
the motion other than energy.$^{\ref{bib:fpl9}}$ 

The reduced action, Eq.\ (\ref{eq:ra}), may be expressed for $V=0$
by

\begin{equation}
W = \hbar \arctan \left(\frac{b \tan [(2mE)^{1/2}x/\hbar ] +
c/2}{(ab-c^2/4)^{1/2}} \right).
\label{eq:raa}
\end{equation}

\noindent For $a=b$ and $c=0$, then $W = (2mE)^{1/2}x$ which
coincides with the classical reduced action.  For $a \ne b$ or $c
\ne 0$, a stratagem is used.  We investigate the $\lim_{E \to \infty} 
(W^{\mbox{\scriptsize quantum}}/W^{\mbox{\scriptsize classical}})$ by using 
l'H\^opital's rule with regard to $E$ to render

\begin{eqnarray*}
\lim_{E\to \infty} \frac{W^{\mbox{\scriptsize quantum}}}{W^{\mbox{\scriptsize 
classical}}} & = & \lim_{E\to \infty} \frac{\hbar \arctan \left(\frac{b \tan 
[(2mE)^{1/2}x/\hbar ] + c/2}{(ab-c^2/4)^{1/2}} \right)}{(2mE)^{1/2}x}\\
& = & \frac{(ab-c^2/4)^{1/2}} {a+b+ (a^2-2ab+b^2-c^2)^{1/2}\cos \{ 
[(2mE)^{1/2}x/\hbar] + \cot ^{-1}[c/(a-b)]\}}
\end{eqnarray*}

\noindent where we have temporarily used the notation of $W^{\mbox{\scriptsize 
quantum}}$ for explicitness. As $W^{\mbox{\scriptsize classical}} = (2mE)^{1/2}$ 
for $E \geq 0$, then we may construct $W$ for energy that is not bound as 

\begin{equation}
\lim_{E\to \infty} W = \frac{(ab-c^2/4)^{1/2} (2mE)^{1/2}} {a+b+ (a^2-2ab+b^2-
c^2)^{1/2}\cos \{ [(2mE)^{1/2}x/\hbar] + \cot ^{-1}[c/(a-b)]\}}.
\label{eq:limqcm}
\end{equation}

\noindent  The quantum reduced action in the limit $E \to \infty$, Eq.\ 
(\ref{eq:limqcm}), manifests the same characteristic residual indeterminacy for 
either $a \ne b$ or $c \ne 0$ that was also manifested for the reduced action in 
Planck limit ($\hbar \to 0$).$^{\ref{bib:f2000}}$  The $E^{1/2}$ factor in the 
argument of the cosine term in the denominator on the right side of Eq.\ 
(\ref{eq:limqcm}) induces in the high energy limit an essential singularity in 
the cosine term. This essential singularity in turn induces an indeterminacy in 
the reduced action, W, in the high energy limit if $a \ne b$ or $c
\ne 0$.  The magnitude of this indeterminacy is a function of the
particular microstate as determined by the set of coefficients
$(a,b,c)$.  The phase shift of this indeterminacy is a function of
$c/(a-b)$.  This indefiniteness in $\lim_{E \to \infty} W$ does not
exist when $E$ is finite.  We shall return latter
in this section after we have established the equations of motion
to show that this indefiniteness can be removed and that another
generator of the motion, Hamilton's principal function, $S$, for
quantum motion in the high energy limit goes to the Hamilton's
principal function, $S^{\mbox{\scriptsize classical}}$, of
classical mechanics. 

We note that for $a=b$ and $c=0$, then $\lim_{E \to \infty} W = W^{\mbox 
{\scriptsize classical}}$ as did $W$ for Planck 
correspondence.$^{\ref{bib:f2000}}$  The quantum reduced action in the high 
energy limit mimics the quantum reduced action for Planck correspondence for all 
microstates ($a,b,c$) in the quantum allowed region because the wave number in 
the cosine term in the denominator has the same form, $(2mE)^{1/2}/\hbar $, 
which manifests the same essential singularity.  Whether $E$ grows without bound 
or $\hbar \to 0$, the wavelength becomes infinitesimally short. Hence both Bohr 
and Planck correspondences in their respective limits manifest an equivalent 
indetermination for the quantum reduced action in the classically allowed 
region.  This allows us to extract the results for Planck 
correspondence$^{\ref{bib:f2000}}$ in the allowed region for our investigation 
of Bohr correspondence. 
 
The quantum conjugate momentum for $V=0$ can be expressed as

\begin{equation}
W_x=\frac{2(2mE)^{1/2}(ab-c^2/4)^{1/2}}{a+b+[(a-b)^2+c^2]^{1/2}\cos
\{[2(2mE)^{1/2}x/\hbar ]+\cot ^{-1}[c/(a-b)]\}}.
\label{eq:qcm}
\end{equation}

\noindent In the limit $E \to \infty$ the cosine term in the
denominator in Eq.\ (\ref{eq:qcm}) fluctuates with an
infinitesimally short wavelength.  This induces in $W_x$ a
residual indeterminacy in the high energy limit.  In the trajectory
representation, we know the microstates, but in the Schr\"odinger representation 
we do not know the microstate because the Schr\"odinger representation assumes a 
reduced set of initial values insufficient to specify the
microstate. For finite $E$, $W_x$ is always specified by $(E,a,b,c,x)$ in the
trajectory representation, but the Copenhagen interpretation denies
knowledge of $(a,b,c)$ while championing Heisenberg uncertainty.  

Nevertheless, we can evaluate its average momentum by averaging
$W_x$ over one cycle of the cosine term.$^{\ref{bib:f2000}}$  We shall use this 
averaging process to
gain insight into the quantum term $\hbar ^2\langle W;x \rangle
/(4m)$.  Nothing herein implies that we are considering an ensemble
of identical microstates rather than a solitary microstate.  The
averaging process leads to$^{\ref{bib:f2000}}$

\begin{eqnarray}
\Bigl\langle \lim_{E \to \infty} W_x\Bigr\rangle _{\mbox{\scriptsize
ave}} & = & \lim_{E \to \infty} \frac{(2mE)^{1/2}}{\hbar \pi }
\int_{\frac{-\hbar \pi }{(8mE)^{1/2}}}^{\frac{\hbar \pi
}{(8mE)^{1/2}}} W_x(E,a,b,c,x+x')\, dx' \nonumber \\
& = & \frac{2(2mE)^{1/2}(ab-c^2/4)^{1/2}}{(a+b)[1-\frac{(a-b)^2-c^2}{(a
+b)^2}]^{1/2}} = (2mE)^{1/2}.
\label{eq:aqcm}
\end{eqnarray}

\noindent [We have changed for this step the operational order of
evaluating the high energy limit and averaging on the right side of
Eq.\ (\ref{eq:aqcm}) because the averaging domain is dependent upon
$E$.  We continue this practice throughout.]  In the high energy limit, the 
average conjugate momentum is the classical momentum and
microstate information as specified by the set of coefficients
$(a,b,c)$ is lost.

The average for $\lim_{E \to \infty}W_x^2$ is given as$^{\ref{bib:f2000}}$ 

\begin{equation}
\Bigl\langle\lim_{E \to \infty}W_x^2\Bigr\rangle_{\mbox{\scriptsize
ave}} = \Bigl\langle\Bigl(\lim_{E \to
\infty}W_x\Bigr)^2\Bigr\rangle_{\mbox{\scriptsize ave}} = mE
\frac{a+b}{(ab-c^2/4)^{1/2}} \geq 2mE.
\label{eq:aqcm2}
\end{equation}  

\noindent  If we identify $W_x^2/(2m)$ as the effective kinetic
energy, then the average of the high energy limit of the effective
kinetic energy for $V=0$ is greater than $E$ for $a \ne b$ or $c
\ne 0$ and is equal to $E$ for $a=b$ and $c=0$. 

Now let us examine the variance of $W_x$ in the high energy limit. 
By Eqs.\ (\ref{eq:aqcm}) and (\ref{eq:aqcm2}), we have$^{\ref{bib:f2000}}$

\begin{equation}
\Bigl\langle \lim_{E \to \infty} W_x^2\Bigr\rangle
_{\mbox{\scriptsize ave}} - \Bigl( \Bigl\langle \lim_{E \to \infty} 
W_x\Bigr\rangle _{\mbox{\scriptsize ave}} \Bigr)^2 =
2mE\frac{a+b-(4ab-c^2)^{1/2}}{(4ab-c^2)^{1/2}} \geq 0.
\label{eq:vqcm}
\end{equation}

\noindent Even in the high energy limit, the variance of the quantum
conjugate momentum, $W_x$, still is a function of the coefficients
$(a,b,c)$ that, in turn, manifest microstates.  For $a=b$ and
$c=0$, then the variance of $W_x$ is zero.  In this particular
microstate, the quantum motion reduces to classical motion for any
value of $E$ because the additional necessary initial values of
the QSHJE, $[W_x(x_o),W_{xx}(x_o)]$ are both zero for a given
energy $E$.   
 
The average energy associated with the Schwarzian derivative term
of the QSHJE in the high energy limit is given from Eqs.\
(\ref{eq:aqcm}--\ref{eq:vqcm}) by$^{\ref{bib:f2000}}$ 

\begin{equation}
\Bigl\langle \lim_{E \to \infty} \frac{\hbar ^2}{4m} \langle
W;x\rangle \Bigr\rangle _{\mbox{\scriptsize ave}} =
E\left(1-\frac{(a+b)/2}{(ab-c^2/4)^{1/2}}\right) =
-\frac{\mbox{variance\ of\ }{\textstyle \lim_{E \to \infty}}W_x}{2m}
\le 0.
\label{eq:bqp}
\end{equation}

\noindent  So the average energy, in the high energy limit, of the
quantum term, $\hbar ^2\langle W;x\rangle /(4m)$, which is also
known for unbound states as Bohm's quantum potential, $Q$, is
proportional to the negative of the variance of the high energy limit
of the conjugate momentum.  The quantum potential is a function of
the particular microstate and may be finite even in the high energy
limit as shown by Eq.\ (\ref{eq:bqp}).  As such, this potential is
not a function of spatial position alone but is path dependent and,
thus, cannot be a conservative potential.  

Let us now consider the equation of motion that is given by
Jacobi's theorem, $t-t_o=W_E$.  For the free particle with energy
$E$, the motion is given by$^{\ref{bib:f2000}}$

\begin{equation}
t-t_o=\frac{(ab-c^2/4)^{1/2}(2m/E)^{1/2}x}{a+b+(a^2-2ab+b^2+c^2)^
{1/2}\cos \{2(2mE)^{1/2}x/\hbar +\cot ^{-1}[c/(a-b)]\}}.
\label{eq:qeom}
\end{equation}

\noindent For $a=b$ and $c=0$, Eq.\ (\ref{eq:qeom}) reduces to the classical 
equation of motion, $t-t_o=(2mE)^{1/2}x$.

Let us now evaluate Hamilton's principal function, $S=W-Et$, in the
high energy limit.  Since we have solved the equations of motion, we
are able to show that $S$ is the time integral of the Lagrangian of
classical mechanics, $L^{\mbox{\scriptsize classical}}$.  For
$V=0$, the classical Lagrangian is a constant give by
$L^{\mbox{\scriptsize classical}}=E$.  As the right sides of Eqs.\
(\ref{eq:limqcm}) and (\ref{eq:qeom}) are dynamically similar except for the 
factor $E$, the reduced action in the
high energy limit may be expressed as a function of time by
$\lim_{E \to \infty} W=2E(t-t_o)$.  The high energy limit of the
reduced action is independent of the set of coefficients $(a,b,c)$. 
Subsequently, Hamilton's principal function may be expressed 
by$^{\ref{bib:f2000}}$

\begin{equation}
\lim_{E \to \infty} S = E(t-t_o) = \int_{t_o}^t L^{\mbox{\scriptsize
classical}}\, dt = S^{\mbox{\scriptsize classical}}
\label{eq:hpft}
\end{equation}

\noindent independent of microstate, $(a,b,c)$.

Let us now return to the equation of motion.  As before with Eq.\
(\ref{eq:qcm}), in the limit $E \to \infty$ the cosine term in the
denominator on the right side of Eq.\ (\ref{eq:qeom}) fluctuates with an 
infinitesimally short wave
length.  This induces a residual indeterminacy in $t(x)$.  Again as
before in the classical limit, we can evaluate the average (this
time for $t$ rather than for $W_x$) by averaging the right side of
the preceding equation over one cycle of the cosine term in the
denominator although here we have $x$ as a factor in the numerator. 
As in our investigation of Planck correspondence,$^{\ref{bib:f2000}}$ the $x$
factor in the numerator is fasted to a single value over the
infinitesimally short wavelength of the cosine term.  This leads to

\begin{equation}
\Bigl\langle \lim_{E \to \infty} (t-t_o)\Bigr\rangle
_{\mbox{\scriptsize ave}} =
\frac{[(ab-c^2/4)(2m/E)]^{1/2}x}{(a+b)[1-\frac{(a-b)^2-c^2}{(a+b)
^2}]^{1/2}} = \left(\frac{m}{2E}\right)^{1/2}x
\label{eq:aqeom}
\end{equation} 

\noindent independent of the coefficients $(a,b,c)$.  While
individual microstates have a residual indeterminacy in the
high energy limit in its motion in the $[x,t]$ domain, this
indeterminacy is centered about the classical motion in the $[x,t]$
domain regardless of which particular microstate is specified. 
Nevertheless, the degree of indeterminacy is a function of the
microstate as specified by $(a,b,c)$.

\section{CLASSICALLY FORBIDDEN CASE}  
The QSHJE renders real solutions, including conjugate momentum, for the 
trajectory even in the classically forbidden region.  On the other hand, the 
CSHJE has a turning point at the WKB turning point.  If classical trajectories
were permitted beyond the turning point, the classical momentum
would become imaginary.  Let us now examine a particle in the
classically forbidden region, $E<V(x)$.  Thus, we have to modify Bohr 
correspondence in the classical forbidden region to permit $E$ to grow to less 
than the upper bound of $V(x)$ for $x$ beyond the WKB turning point.  (Note that 
this condition permits tunneling, which is not examined herein but has been 
examined for sub-barrier energies well below the barrier potential in Ref.\ 
\ref{bib:afb20}.)  For this investigation, we assume a finite potential well in 
the classically forbidden regions that $V \gg 1$ but remains finite for finite 
$x$.  We investigate Bohr correspondence in the classically forbidden region by 
assuming that the $V-E$ is finite and positive.  Here, we choose to examine a 
square well given by

\[
V= \left\{ \begin{array}{lc}
               U>E\gg 1, &  \ |x| \ge q \\
               0,      &  \ |x|< q.
           \end{array}
     \right.
\]

\noindent The trajectories for the bound states for this square well have 
already been studied.$^{\ref{bib:fpl13}}$ For brevity and without any loss of 
physics, we only examine the symmetric bound states herein. The set of 
independent solutions $(\phi ,\theta )$ for
this square well is chosen such that $\phi $ represents the symmetric bound 
state
given by

\begin{equation}
\phi  =  \left(\frac{2m}{\hbar ^2k^2(ab-c^2/4)}\right)^{1/4} \cdot \left\{
\begin{array}{lr}
                   \cos (kq) \exp [-\kappa (x-q)], & x>q \\ [.08 in]
                   \cos (kx), & -q \leq x \leq q \\  [.08 in]
                   \cos (kq) \exp [\kappa (x+q)] & x<-q
                   \end{array}
             \right.
\label{eq:phisw}
\end{equation}

\noindent where $k=(2mE)^{1/2}/\hbar $ and $\kappa = [2m(U-E)]^{1/2}/\hbar $.  
The other solution, $\theta $, is unbound and is not unique as any
amount of $\phi $ may be added to it.  While $\phi $ is symmetric for the
symmetric bound state, the corresponding $\theta $ that we have chosen is
antisymmetric. We present this unbound solution as

\begin{equation}
\theta =  \left(\frac{2m}{\hbar ^2k^2(ab-c^2/4)}\right)^{1/4} \cdot \left\{
\begin{array}{lr}
\displaystyle{\frac{\exp [\kappa (x-q)] - \cos (2kq) \exp [-\kappa
(x+q)]}{2\sin (kq)}}, & x>q \\ [.08 in]
\sin (kx),  & -q \leq x \leq q \\ [.08 in]
\displaystyle{\frac{\cos (2kq) \exp [\kappa (x+q)] - \exp [-\kappa
(x+q)]}{2\sin (kq)}}, & x<-q. \\
\end{array}
\right.
\label{eq:thetasw}
\end{equation}

\noindent The corresponding Wronskian obeys ${\cal W}^2(\phi ,\theta ) =
2m/[\hbar ^2(ab-c^2/4)] > 0$ as expected.  For bound states, microstates of
the Schr\"{o}dinger wave function exist where the particular choice of the set
of coefficients $(a,b,c)$ specifies the microstate and a unique trajectory in 
phase space for a given quantized energy $E$.$^{\ref{bib:fpl9}}$  

The dwell time, $t_{\pm R}$, is the time particle spends in the a classically 
forbidden region during the round trip transit along its trajectory from the 
wall of the square well at $x=\pm q$ out to its reflection at the turning point 
at $x=\pm \infty $ and then its return back to $x=\pm q$.  The dwell time, 
$t_R$, has been shown to be$^{\ref{bib:fpl13}}$  

\begin{equation}
t_{\pm R} = 2 \frac{(ab-c^2/4)^{1/2}[1+(\kappa /k)^2]}{a \pm c(\kappa /k) + 
b(\kappa /k)^2} \frac{m}{\hbar \kappa k}
\label{eq:reflextime}
\end{equation}

\noindent where the sign for the coefficient $c$ in the denominator is dependent 
upon which interface, $x=\pm q$ is applicable.  (Lest we forget, $c$ itself may 
be negative too.)  The trajectory for the microstate, $(a,b,c)$, will not be 
symmetric if $c\ne 0$.  The existence of unsymmetric microstates of symmetric 
Schr\"odinger wave functions have already been discussed 
elsewhere.$^{\ref{bib:prd26}}$  For $a=b$ and $c=0$, Eq. (\ref{eq:reflextime}) 
becomes 

\[
t_R = t_{+R} = t_{-R} = 2m/(\hbar \kappa k) = \hbar /[E(U-E)]^{1/2}
\]

\noindent which is consistent with findings of Hartman$^{\ref{bib:hartman}}$ and 
Fletcher$^{\ref{bib:fletcher}}$ for tunnelling times, $t_T$ when one takes into 
account that$^{\ref{bib:stoveng},\ref{bib:or}}$ $t_T = t_R$.  Note that $t_R$ is 
inversely proportional to $\kappa $ or $(U-E)^{1/2}$.  This implies that the 
particle velocity increases as $(U-E)^{1/2}$ increases which is consistent with 
the findings of Barton.$^{\ref{bib:ap166}}$  This seems to be counterintuitive 
and to permit superluminal velocities, and much ado has been reported about this 
aspect of nonlocality complete with {\it Alice in Wonderland} 
cartoons.$^{\ref{bib:chiao}}$  In the last paragraph of this section, we shall 
discuss this aspect of nonlocality further.  

The period of libration for the trajectory for the corresponding microstate is 
given as$^{\ref{bib:fpl13}}$  

\begin{equation}
t_{\mbox{\scriptsize{libration}}}=4 \frac{(ab-c^2/4)^{1/2} [1+(\kappa /k)^2]
[a+b(\kappa /k)^2]} {a^2 + (2ab-c^2)(\kappa /k)^2 + b^2(\kappa /k)^4}
\frac{m(q+\kappa ^{-1})}{\hbar k}.
\label{eq:lte}
\end{equation}

\noindent The fractional time of the libration period that the particle spends 
in the classically forbidden region is given by$^{\ref{bib:f2000}}$

\begin{equation}
\frac{t_{+R} + t_{-R}}{t_{\mbox{libration}}}=\frac{1}{\kappa q + 1} = 
\frac{\hbar }{\hbar + [2m(U-E)]^{1/2}q}.
\label{eq:fractime}
\end{equation}

\noindent The fractional time is independent of the coefficients $(a,b,c)$.  
Also as $E$ approaches $U$ from below, the fractional time approaches one.  For 
Planck correspondence $(\hbar \to 0)$, a previous investigation showed that the 
particle does not spend any time in the classically forbidden 
region.$^{\ref{bib:f2000}}$  This is also confirmed by Eq.\ (\ref{eq:fractime}) 
as its right side goes to zero as $\hbar \to 0$.  While Bohr correspondence in 
the trajectory representation in the classically allowed region is analogous to 
Planck correspondence, they differ in the classically forbidden region. 

Let us return to the allegedly anomaly of the counterintuitive nature that dwell 
times for reflection, $t_R$, decrease with increasing value of $(U-E)^{1/2}.$  
When examined from a  dwell time aspect, this anomaly that manifests nonlocality 
is reasonable.  But when fractional time in the classically forbidden region, 
Eq.\ (\ref{eq:fractime}), is considered, this nonlocal system is no longer 
counterintuitive.  The particle, as manifested by Eq.\ (\ref{eq:fractime}), 
spends a greater proportion of its time in the classically forbidden region as 
$E$ appproaches $U$ from below.

\section{WKB TURNING POINT}
Let us now investigate the trajectory in
the high energy limit in the vicinity of the WKB turning point.  Here, too, we 
require that the value $V(x)$ be in the neighborhood of $E$ even though we let 
$E$ get large without limit.  We also examine the transition between the 
classically allowed and classically forbidden regions in the high energy limit 
as the trajectory transits the WKB turning point.  We choose the potential
to be

\begin{equation}
V=fx,
\label{eq:vtp}
\end{equation}

\noindent which represents a constant force $f>0$ acting on our
particle.  Any well-behaved one-dimensional potential for which
the force remains finite and continuous can always be approximated
by a linear potential in a sufficiently small region containing the
WKB turning point as an interior point.  As we shall show, while the value of 
$f$ must be matched to the force exerted on the particle by the potential at the 
WKB turning point, it is not necessarily a global value for this investigation.

Let us digress briefly.  In the previous two sections, we examined
potentials that were at least piecewise constant.  Even though the
independent solution set, $(\phi ,\theta )$, for a step potential
is mathematical simpler, such a potential does not have a classical
short-wave correspondence at the WKB turning point for the relative
change in the potential over a wavelength remains large
there.$^{\ref{bib:ajp60}}$   

For a particle with energy $E$, the WKB turning point, $x_t$, is
given by $x_t=E/f$.  An acceptable set of independent solutions,
$(\phi ,\theta )$, to the Schr\"{o}dinger equation is formed  
of Airy functions given by

\begin{equation}
\phi = \frac{(2m)^{1/12} \pi^{1/2} \mbox{Ai}[(2mf/\hbar
^2)^{1/3}(x-E/f)]}{(\hbar f)^{1/6}(ab-c^2/4)^{1/2}} \ \ \ \
\mbox{and} \ \ \ \ \theta = \frac{(2m)^{1/12} \pi^{1/2}
\mbox{Bi}[(2mf/\hbar ^2)^{1/3}(x-E/f)]}{(\hbar f)^{1/6}(ab-
c^2/4)^{1/2}}.
\label{eq:isl}
\end{equation}

The reduced action for the linear potential specified by Eq.\
(\ref{eq:vtp}) is given by

\begin{equation}
W = \hbar \arctan \left(\frac{b \mbox{Bi}(\xi /\hbar
^{2/3})/\mbox{Ai}(\xi /\hbar ^{2/3}) + c/2}{(ab-c^2)^{1/2}}\right)
\label{eq:qhcft}
\end{equation}

\noindent where $\xi=(2mf)^{1/3}(x-e/f)$.

We examine the quantum equation of motion that is given by
Jacobi's theorem.  For the particle with energy $E$ and subject to
the linear potential, Eq.\ (\ref{eq:vtp}), the motion is given by

\begin{equation}
t-t_o=\frac{\hbar ^{1/3}}{\pi }\frac{(ab-
c^2/4)^{1/2}(2m/f^2)^{1/3}}{a \mbox{Ai}^2(\xi /\hbar ^{2/3}) + b
\mbox{Bi}^2(\xi /\hbar ^{2/3}) + c \mbox{Ai}(\xi /\hbar ^{2/3})
\mbox{Bi}(\xi /\hbar ^{2/3})}.
\label{eq:qeomtp}
\end{equation}

\noindent As $E$ increases in going to the high energy limit, the
WKB turning point moves out continuously where, for nonlinear potentials, the 
instantaneous value of $f$ must be adjusted continuously to manifest the 
gradient of the potential in the current vicinity of the turning point.  Taking 
the high energy limit does not change the character of the physics of the 
situation.  About $x_t = E/f$ or $\xi = 0$, there will be a finite neighborhood, 
part in the classically allowed region and part in the classically forbidden 
region, in which $(E-fx)$ or $\xi $ will be small.  In this finite neighborhood, 
the fractional change in value of $(E-V)$ will be large over a wavelength which 
will induce quantum effects.  

For $a=b$ and $c=0$, the motion of the particle displaced form the turning point 
is easily analyzed.  For classically allowed $x$ sufficiently displaced from the 
turning point at $E/f$ with $a=b$ and $c=0$, the Airy functions may be 
represented by asymptotic approximations that will manifest in the classically 
allowed region the classical motion given by$^{\ref{bib:f2000}}$

\begin{equation}
t-t_o \bigg|_{a=b;c=0} = \frac{[2m(E-fx)]^{1/2}}{f}, \ \ E-fx \gg 1.
\label{eq:eomfx}
\end{equation}

\noindent Likewise, for $x$ displaced from $x_t$ into the classically forbidden 
region, the velocity has been shown elsewhere to grow without bound as the 
particle penetrates further into the classically forbidden 
region.$^{\ref{bib:f2000}}$ 

While the Planck correspondence, $\hbar \to 0$, showed that the measure of the  
transitional neighborhood across a WKB turning point reduces to 
zero,$^{\ref{bib:f2000}}$ the corresponding transitional neighborhood in the 
high energy limit remains finite.  Nevertheless, for bound states in the high 
energy limit, the width the classically allowed region for many potentials may 
grow without bound.  In such case, the relative measure of the transitional 
domain to the classically allowed domain may become infinitesimally small. In 
this sense, Bohr correspondence and Planck correspondence are partially similar.  

\section{INITIAL VALUES}  
Basically, our investigation of the Bohr correspondence supports our findings 
for Planck correspondence regarding initial conditions.  The QSHJE is a third 
order nonlinear
differential equation while the CSHJE is first order.  The reduced
action, $W$ or $W^{\mbox{\scriptsize classical}}$, does not
explicitly appear in either the quantum or classical equation
respectively.  Then, a set of necessary and sufficient initial
values at $x_o$ needed to specify the quantum conjugate momentum in
addition to the constant of motion $E$ is $[W_x(x_o),W_{xx}(x_o)]$
while just $E$ is necessary and sufficient to specify the classical
conjugate momentum.  Subsequently, the quantum and
classical reduced actions are known to within an arbitrary
integration constant.  The additional initial condition $W(x_o)$ specifies the 
integration constant $K$ in Eq.\ (\ref{eq:ra}).

The initial values for the solution of the QSHJE specify the
particular microstate.$^{\ref{bib:prd34},\ref{bib:fpl9}}$  We note
that when $a=b$ and $c=0$, the quantum trajectory solutions in the high energy 
limit for $V=0$ and for $V=fx$, with $x$ sufficiently inside the classically 
allowed region, reduced to the classical trajectory solutions as shown by Eqs.\ 
(\ref{eq:qeom}) and (\ref{eq:eomfx}) respectively.  The underlying physics for 
$V=0$ is that, for coefficients $a=b$ and $c=0$, $W_x$ by Eq.\
(\ref{eq:qcm}) becomes a constant, $(2mE)^{1/2}$, consistent with the 
corresponding $W_x^{\mbox{\scriptsize classical}}$.  Even for finite energy,
selecting $a=b$ and $c=0$ causes $W_{xx}$ and $W_{xxx}$ to be zero
consistent with classical mechanics for $V=0$.  If $E$ is unknown,
then $[W_x(x_o),W_{xx}(x_o),W_{xxx}(x_o)]$ forms a necessary and
sufficient set of initial values to specify
$W(x)$.$^{\ref{bib:prd29}}$  For the linear potential $V=fx$,
selecting $a=b$ and $c=0$, we have by Eq.\ (\ref{eq:eomfx}) classical motion in 
the classically allowed region well displaced from the turning point.  Thus,
choosing $a=b$ and $c=0$ tacitly induces the necessary initial
values $[W_x(x_o),W_{xx}(x_o)]$ for the QSHJE to correspond to the
superfluous initial values $[W_x^{\mbox{\scriptsize
classical}}(x_o),W_{xx}^{\mbox{\scriptsize classical}}(x_o)]$ for
the corresponding CSHJE.  Another view is that classical mechanics
inherently assumes that $a=b$ and $c=0$ for our choices, Eqs.
(\ref{eq:isc}) and (\ref{eq:isl}), for the set $(\phi ,\theta )$ of
independent solutions.  For completeness, had we chosen a different
set of independent solutions than those specified by Eqs.\
(\ref{eq:isc}) or (\ref{eq:isl}) for $V=0$ and $V=fx$ respectively,
then, for that potential, we would have had to choose a different
set of coefficients $(a,b,c)$ to achieve Bohr correspondence to
classical mechanics.

\section{LOSS OF INFORMATION}  Here too, our investigation of the Bohr 
correspondence supports our prior findings regarding Planck 
cor\-res\-pon\-dence.$^{\ref{bib:f2000}}$  Passing to the high energy limit
incurs a loss of information, associated with a set of necessary
and sufficient initial values, due to going from the third-order
QSHJE to the first-order CSHJE.  Likewise, even for finite energy, the 
Copenhagen interpretation looses the information inherent to
microstates of the trajectory representation despite the Copenhagen
school asserting that the Schr\"{o}dinger wave function is
exhaustive.  For completeness, we study loss of information for
both cases and compare our results with
't~Hooft.$^{\ref{bib:thooft}}$  There are similarities and
significant differences.

First, we shall examine the quantum level ($E$ finite).  As
already noted, the trajectory representation in its Hamilton-Jacobi
formulation manifests microstates not discernible by the
Schr\"{o}dinger representation showing that the Schr\"{o}dinger
wave function cannot be the exhaustive description of
nature.$^{\ref{bib:pla249},\ref{bib:fpl9},\ref{bib:prd26}}$  For a
specified energy $E$, each microstate, by itself, specifies the
Schr\"{o}dinger wave function.  Yet, each microstate of energy
eigenvalue $E$ has a distinct trajectory specified by the set of
initial values $[W_x(x_o),W_{xx}(x_o)]$.  Are these microstates
primordial at the Planck level?  Yes, we have already shown
elsewhere$^{\ref{bib:prd29}}$ that the trajectories are
deterministic and that the trajectory representation has arbitrary
initial conditions at
the Planck level.  As the Copenhagen school asserts that $\psi $
should be the exhaustive description of natural phenomenon, the
Copenhagen school denies that primordial microstates could exist. 
Viewed externally, the Copenhagen school unwittingly makes $\psi $
to be a de facto equivalence class of any putative microstates. 
This is consistent with the QSHJE being more fundamental than the
stationary Schr\"{o}dinger equation, in contrast to Messiah's
assertation,$^{\ref{bib:messiah}}$ because, as shown
elsewhere,$^{\ref{bib:fpl9}}$ the bound-state boundary conditions
of the QSHJE do not generate a unique solution but rather generate
an infinite number of ``primordial" microstates while the boundary
conditions for the Schr\"{o}dinger equation do generate a unique
``equivalence-class" bound-state wave function.  Let us make a few
comparisons with a 't~Hooft process.  The primordial microstates
are deterministic trajectories of discrete energy $E$ in contrast
to the 't~Hooft primordial states that are of the continuum. 
Nevertheless, as all initial values are allowed for the
microstates,$^{\ref{bib:prd29}}$ the trajectories for any
equivalence class manifested by $\psi $ densely spans finite phase
space.  Also, the Copenhagen school looses information on
primordial microstates by default and not through a 't~Hooft
dissipative process.      

Second, let us examine loss of information in the trajectory
representation by executing the high energy limit. In the trajectory
representation, residual indeterminacy manifests loss of
information.  The residual indeterminacy for the solution, $W_x$,
of the QSHJE in the high energy limit is given for $V=0$ by the trigonometric 
terms

\[
[(a-b)^2+c^2]^{1/2}\cos \{[2(2mE)^{1/2}x/\hbar]+\cot ^{-1}[c/(a-
b)]\}
\]

\noindent in the denominator of Eq.\ (\ref{eq:qcm}).  Here, we set the square of 
the
amplitude of these trigonometric terms to be $A = (a-b)^2+c^2$
where A is dimensionless.  The
phase shift of the argument of these trigonometric terms is
manifested by $\cot ^{-1}[c/(a-b)]$.  The factor $E$ in 
the argument of these trigonometric terms induces an indeterminacy
in the high energy limit that makes the phase shift due to $\cot ^{-
1}[c/(a-b)]$ irrelevant.  This represents a loss of information. 
On the other hand, the phase shift, $\cot ^{-1}[c/(a-b)]$, is not
redundant for specifying the set coefficients $(a,b,c)$ from the
set of necessary and sufficient initial values
$[W_x(x_o),W_{xx}(x_o)]$ for the QSHJE for a finite $E$.  Hence, the
irrelevance of the phase shift gives the set of coefficients
$(a,b,c)$ another degree of freedom that makes the set of
coefficients underspecified in the high energy limit.  This
underspecification of coefficients $(a,b,c)$ permits the primordial
microstates to form into equivalence classes where the primordial
microstates establish the membership within any particular
equivalence class and become identical with one another in the
high energy limit.  This information loss differs with that of
't~Hooft.  As before, the primordial microstates have discrete
rather than continuum energies.  Also again, no dissipation of
information occurs in the trajectory representation when going to
the high energy limit, but rather this loss of information induces an
indeterminacy.  

We may generalize to say that as classical mechanics has a smaller
set of necessary and sufficient initial values than the trajectory
representation of quantum mechanics, then there is some loss of
information and the formation of equivalence classes as we go to
the high energy limit.  Also, the Copenhagen school by precept
considers $\psi $ to be exhaustive and disregards any microstate
information.  In either case, this loss of information is not due
to any dissipation as it is in 't~Hooft's proposal.  Without
dissipation, this loss of information may occur for 
the quantum stationary Hamilton-Jacobi process in the high energy limit. 
      
\section{INDETERMINACY}

Let us begin by contrasting the residual indeterminacy of the trajectory 
representation to Heisenberg uncertainty.  Our findings regarding the 
relationship between residual indeterminacy and the Heisenburg uncertainty 
principle in the high energy limit are consistent with our analogous findings 
for $\hbar \to 0$.$^{\ref{bib:f2000}}$  Residual indeterminacy exists when 
energy increases without bound $(E \to \infty)$.  For finite $E$, there is no 
indeterminacy as the trajectory and $E$ are specified by the sufficient set 
initial conditions $(\dddot{x}_o,\ddot{x}_o,\dot{x}_o,x_o)$ or 
$[W_{xxx}(x_o),W_{xx}(x_o),W_x(x_o),x_o]$.$^{\ref{bib:prd29}}$  On the other 
hand, Heisenberg uncertainty exists for finite $E$.  We note that residual 
indeterminacy is consistent with the findings of Faraggi and Matone that the
equivalence principle exists for quantum mechanics but not for
classical mechanics.$^{\ref{bib:fm3}}$  Otherwise, the trajectory
representation remains causal$^{\ref{bib:ijmpa14}}$ and
deterministic.  

Heisenberg uncertainty exists in the $[x,p]$ domain (where $p$ is
momentum) since the Hamiltonian operates in the $[x,p]$ domain.  On
the other hand, the trajectory representation through a canonical
transformation to its Hamilton-Jacobi formulation operates in the
$[x,t]$ domain.$^{\ref{bib:park}}$  Residual indeterminacy of the
trajectory representation is in the $[x,t]$ domain, cf. Eqs.\
(\ref{eq:aqeom}) and (\ref{eq:eomfx}).

The Copenhagen representation, based upon Heisenberg uncertainty tacitly, 
uses an insufficient set of initial values to specify quantum motion.  The 
subset of initial values that the Copenhagen representation is sufficient to 
specify $\psi $ but precludes knowledge of its microstates. The
Heisenberg uncertainty principle masks the fundamental cause of
uncertainty in the Copenhagen interpretation.  As long as the
Copenhagen interpretation uses an insufficient set of initial
values to solve the QSHJE, it cannot achieve certainty. Carroll has speculated 
that the Copenhagen substitute for exact microstate knowledge is, in itself, an 
uncertainty principle.$^{\ref{bib:carroll2}}$  Carroll goes on to note that 
standard quantum mechanics can still be used with the acknowledgement that 
microstate knowledge has been sacrificed (i.e., quantum mechanics in Hilbert 
space is imprecise by construction so a probabilistic theory 
follows).$^{\ref{bib:carroll2}}$

In closing, we remark on the impact of residual indeterminacy.  Our findings for 
the trajectory representation in the high energy limit are consistent with 
Liboff's findings$^{\ref{bib:liboff}}$ for the Schr\"odinger representation.  In
the high energy limit, $E \to \infty$, the trajectory representation
of quantum mechanics does not generally go to classical mechanics
in contradiction to Bohr's correspondence principle.  Nor does it go to
statistical mechanics as the amplitude of the indeterminacy is
given by $[(a-b)^2+c^2]^{1/2}$ for the sets of independent
solutions of the Schr\"{o}dinger equation used herein, cf. Eqs (\ref{eq:isc}) 
and (\ref{eq:isl}). 
 
\begin{center}
{\bf Acknowledgement}
\end{center}

I heartily thank M. Matone, A. E. Faraggi, and R. Carroll for their
stimulating discussions and encouragement.  I also thank G.
't~Hooft for discussing his ideas on information loss and
equivalence classes.

\paragraph{References}
\begin{enumerate}\itemsep -.06in
\item \label{bib:liboff} R.\ L.\ Liboff, Phys.\ Today {\bf 47}(2) 55 (1984); 
Ann.\ Fond.\ L.\ deBroglie {\bf 5}, 215 (1980); Int.\ J.\ Theor.\ Phys.\ {\bf 
18}, 185 (1979); Fond.\ Phys.\ {\bf 5}, 271 (1975).
\item \label{bib:ajp60} J.\ Ford and G.\ Mantica,  Am.\ J.\
Phys.\ {\bf 60}, 1086 (1992).
\item \label{bib:f2000} E.\ R.\ Floyd, Int. J. Mod. Phys. {\bf A 15}, 1363 
(2000).
\item \label{bib:bohr} N. Bohr in {\it Sources of Quantum Mechanics} edited by 
B. L. Van der Waerden (Dover, New York, 1967) pp. 95-137.
\item \label{bib:rosenfeld} L.\ Rosenfeld, ed.\ {\it Niels Bohr,
Collected Works} Vol.\ 3 (North Holland, New York, 1976)
\item \label{bib:f1988} E.\ R.\ Floyd, Int.\ J.\ Theor.\ Phys. {\bf 27}, 273 
(1988).
\item \label{bib:bfm} G.\ Bertoldi, A.\ E.\ Faraggi and M.\ Matone, Class.\ 
Quant.\ Grav.\ {\bf 17}, 3965 (2000), hep-th/9909201.
\item \label{bib:fm1} A.\ E.\ Faraggi and M.\ Matone,  Phys.\
Lett.\ {\bf B 437}, 369 (1998), hep-th/9711028; 
Phys.\ Lett.\ {\bf B 445}, 357 (1998), hep-th/9809126; Phys.\
Lett.\ {\bf B 450}, 34 (1999), hep-th/9705108.
\item \label{bib:pla249} A.\ E.\ Faraggi and M.\ Matone, Phys.\ Lett.\ {\bf A 
249}, 180 (1998), hep-th/9801033. 
\item \label{bib:fm3} A.\ E.\ Faraggi and M.\ Matone, Phys.\
Lett.\ {\bf B 445}, 77 (1998) hep-th/9809125;  Int.\ J.\ Mod.\ Phys.\ {\bf A 15}, 1869 
(2000), hep-th/9809127.
\item \label{bib:messiah} A.\ Messiah, {\it Quantum Mechanics}
Vol.\ 1 (North Holland, New York, 1961) p.\ 232.
\item \label{bib:prd34} E.\ R.\ Floyd,  Phys.\ Rev.\ {\bf D
34}, 3246 (1986).  
\item \label{bib:afb20} E.\ R.\ Floyd,  An. Fond. Louis de
Broglie {\bf 20}, 263 (1995). 
\item \label{bib:carroll} R.\ Carroll, Can.\ J.\ Phys.\ {\bf 77} 319 (1999), 
quant-ph/9904081
\item \label{bib:carroll2} R.\ Carroll, {\it Quantum Theory, Deformation and 
Integrability} (Elsevier, 2000, Amsterdam) pp. 50-56.
\item \label{bib:thooft} G.\ 't~Hooft, Class.\ Quantum Grav.\ {\bf 16}, 3263 
(1999), gr-qc/9903084
\item \label{bib:fpl9} E.\ R.\ Floyd, Found.\ Phys.\ Lett.\
{\bf 9}, 489 (1996) quant-ph/9707051.
\item \label{bib:fpl13} E.\ R.\ Floyd, Found.\ Phys.\ Lett.\ {\bf 13}, 235 
(2000). 
\item \label{bib:prd26} E.\ R.\ Floyd,  Phys.\ Rev. {\bf D
26}, 1339 (1982). 
\item \label{bib:hartman} T.\ E.\ Hartman, J.\ App.\ Phys.\ {\bf 33}. 3247 
(1962).
\item \label{bib:fletcher} J.\ R.\ Fletcher, J.\ Phys. {\bf C 18}, L55 (1985).
\item \label{bib:stoveng} E.\ H.\ Hauge and J.\ A.\ Stoveng, Rev. Mod. Phys.  
{\bf 61}, 917 (1989).
\item \label{bib:or} V.\ S.\ Olkhovsky and E. Racami, Phys. Rep. {\bf 214}, 339 
(1992).
\item \label{bib:ap166} G.\ Barton,  Ann.\ Phys.\ (NY) {\bf
166}, 322 (1986).
\item \label{bib:chiao} R.\ Y.\ Chiao, P.\ G.\ Kwiat, and A.\ M.\ Steinberg, 
Sci.\ Am., {\bf 269}(2), 52 (Aug. 1993) 
\item \label{bib:prd29} E.\ R.\ Floyd,  Phys.\ Rev.\ {\bf D
29}, 1842 (1984).
\item \label{bib:ijmpa14} E.\ R.\ Floyd,  Int.\ J.\ Mod.\
Phys.\ {\bf A 14}, 1111 (1999), quant-ph/9708026; {\bf 16}, 2447 (2001).
\item \label{bib:park} D.\ Park, {\it Classical Dynamics and Its
Quantum Analogues}, 2nd ed.\ (Springer-Verlag, New York, 1990) p.\
142.
\end{enumerate} 

\end{document}